\documentclass[useAMS,usenatbib]{mn2e}

\usepackage{epsfig}
\usepackage{amsfonts}
\usepackage{hyperref}
\usepackage{multicol}
\usepackage{color}

\title[Photometric AGN classification with Machine Learning methods]{ Photometric classification of emission line galaxies with Machine Learning methods}
\author[Cavuoti et al. 2013]{Stefano Cavuoti$^{1,2}$\thanks{e-mail: stefano.cavuoti@gmail.com}, Massimo Brescia$^{1}$, Raffaele D'Abrusco$^{3}$, Giuseppe
Longo$^{2,4}$, \and Maurizio Paolillo$^{2}$.\\ \\
 1 - INAF-Astronomical Observatory of Capodimonte, via Moiariello 16, I-80131 Napoli, Italy\\
 2 - Department of Physical Sciences, University Federico II, via Cinthia 6, I-80126 Napoli, Italy\\
 3 - Harvard Smithsonian Center for Astrophysics, 02138 Cambridge, MA, USA\\
 4 - Visiting Associate, California Institute of Technology, 91125 Pasadena, CA, USA}
\date{Accepted 2013 October, 10 ; Received 2013 September, 24 ; in original form 2013 August, 1 }
\pagerange{\pageref{firstpage}--\pageref{lastpage}} \pubyear{2013}

\begin{document}

\label{firstpage}
\maketitle

\begin{abstract}

In this paper we discuss an application of machine learning based methods to the identification of candidate AGN from optical survey data and to the
automatic classification of AGNs in broad classes. We applied four different machine learning algorithms, namely the Multi Layer Perceptron (MLP), trained
respectively with the Conjugate Gradient, Scaled Conjugate Gradient and Quasi Newton learning rules, and the Support Vector Machines (SVM), to  tackle
the problem of the classification of emission line galaxies in different classes, mainly AGNs vs non-AGNs, obtained using optical photometry in place of
the diagnostics based on line intensity ratios which are classically used in the literature. Using the same photometric features we discuss also the
behavior of the classifiers on finer AGN classification tasks, namely Seyfert I vs Seyfert II and Seyfert vs LINER. Furthermore we describe the
algorithms employed, the samples of spectroscopically classified galaxies used to train the algorithms, the procedure followed to select the photometric
parameters and the performances of our methods in terms of multiple statistical indicators.  The results of the experiments show that the application of
self adaptive data mining algorithms
trained on spectroscopic data sets and applied to carefully chosen photometric parameters represents a viable alternative to the classical methods that
employ time-consuming spectroscopic observations.
\end{abstract}
\section{Introduction}
Active galactic nuclei (AGNs) and their high redshift counterpart, quasars, are the most luminous long-lived discrete objects in the Universe and are therefore crucial to address a variety of astrophysical and cosmological problems. Individual multi-wavelength studies allow to investigate the physical conditions in the proximity of the central power source, a supermassive black hole with a surrounding accretion disk. The study of well defined samples of AGNs in various environments, both in the local universe and at high redshift, is needed to constrain the various mechanisms invoked to explain galaxy assembly and early evolution \citep{mahajan2010}. It is also needed to explore the role of the environment in triggering or inhibiting nuclear activity \citep{kauffmann2004,popesso2006}.

In spite of the fact that the unified model \citep{antonucci1993}, may provide a unique physical explanation for the central engine and the surrounding
regions, the observable phenomenology of AGNs (and quasars) is quite complex and
encompasses a variety of objects (Seyfert galaxies, liners, quasars, blazars, etc.), which, for the large differences
existing in their observed properties, have for a long time been considered to be different and independent species \citep{antonucci1993}.

This phenomenological complexity is also the reason why there cannot be a unique method equally effective in identifying all AGN
phenomenologies (cf. \citealt{messias2010}) in every redshift range. While it is clear that the most effective and the
most used relies on X-ray emission properties (cf. \citealt{alexander2002}), other types of indicators can be effectively used such as mid-infrared fluxes
(cf. \citealt{donley2007}), color-color diagrams (cf. \citealt{Hatziminaoglou2005}) and even radio data both through peak emission and through unresolved
radio emission \citep{seymour2007}. All these methods have their share of pro \& cons and are more or less biased against the detection of specific
sub-types.

The techniques developed to classify galaxies based on the presence and intensity of spectral emission features have received special attention in the
astronomical literature. Emission lines are visible in the spectra of a large fraction of galaxies of different classes. The relative strengths of multiple
emission lines have been successfully used to infer the class of galaxies observed spectroscopically (see, for example, the ground-breaking work discussed
by~\citealt{veilleux1987}). Among emission line galaxies Seyferts, LINERs and starburst galaxies can be distinguished based on their characteristic
positions in the diagrams generated by the ratios of the equivalent width of the $[OIII](5007)$, $H_{\beta}$ lines on the y axis and $[NII](6584)$,
$H_{\alpha}$ lines on the x axis~\citep{veilleux1987}. In this diagram, starburst galaxies are located in the lower left-hand region,
narrow-line Seyferts are located in the upper right corner and LINERs tend to occupy the lower right-hand
region. Different parametric models of the lines delimiting these different regions have been proposed, based
on the growing availability of large samples of galaxies with spectra~\citep{kewley2001,lamareille2010}. Details
on the most recent parametrization of the line ratio diagnostic diagram are given in Section~\ref{sec:data}.
While such methods based on spectroscopic data provide efficient and reliable classification of
line-emitting galaxies, they are applicable only to galaxies with measured spectra, which usually represent
a small fraction of the total number of galaxies observed in the modern mixed (photometric and spectroscopic)
digital optical surveys (e.~g. the Sloan Digital Sky Survey - SDSS, see~\citealt{york2000}).

In this paper we investigate the possibility to: i) identify candidate AGNs and, ii) classify them in broad classes using optical photometric parameters
only, by means of supervised classification techniques trained on samples of spectroscopically classified galaxies. This study was motivated by the growing
number of planned and ongoing optical surveys covering with high accuracy and depth large portions of the sky, such as KIDS \citep{dejong2013}, DES
\citep{annis2013}, PANSTARRS \citep{tonry2012} etc., where the possibility to identify reliable AGN samples from optical data alone, would be very helpful
in selecting well characterized statistical samples, and to identify candidates for subsequent spectroscopic validation and other follow-up studies.

The new digital surveys produce complex data with many tens or hundreds of parameters measured by automatic methods and carry much more information than in
the past.
This wealth of accurate data has opened the path to the use of data mining methods (typical of machine learning), in place of the usual statistical tools
to perform all sorts of classification, regression and clustering.

An interesting attempt to tackle this problem was performed by \cite{suchkov2005} who tried to reproduce the SDSS classification using only optical colors
and reached the conclusion that \emph{SDSS colors feature prominently in the algorithm used to select AGN candidates for subsequent SDSS spectroscopy}
\citep{suchkov2005}.

The work described here was performed using supervised machine learning methods offered to the community through the DAta Mining \& Exploration Web
Application REsource (DAMEWARE\footnote{\url{http://dame.dsf.unina.it/dameware.html}}). For those who are not familiar with the topic, we shall just
remind that supervised methods learn how to perform a given operation (for instance how to disentangle normal from active galaxies) using a set of well
known examples (also known as \textit{a priori} knowledge or \textit{knowledge base}). The methods we used in this paper are, respectively, Support Vector
Machines (SVM; \citealt{chang2001}) and the Multi Layer Perceptron (MLP) with different types of training rules: the Conjugate gradient (CG;
\citealt{golub1999}), the Scaled Conjugate Gradient (SCG; \citealt{watrous1987}) and the Quasi Newton Algorithm (MLPQNA; \citealt{brescia2012}) and are
shortly summarized in Sec.~\ref{themethods}.

The data set used for the experiments was obtained by joining three catalogues of objects (within the redshift range $0.02<z<0.3$), respectively from
\cite{sorrentino2006}, \cite{kauffmann2003} and \cite{dabrusco2007}, as described in Sec.~\ref{thedata}. The data set contains the photometric parameters
(hereinafter named as input features), as well as flags describing the nature of the objects (in machine learning methods, this flag is also called target)
to be used only in the training and test steps of any experiment. These flags constitute the Knowledge Base (or KB).

We performed three types of experiments. First of all the detection of candidate AGNs (AGN vs non AGN), which is the main experiment; then we tried to
classify Seyfert I against Seyfert II type galaxies; finally we also tested the possibility to distinguish Seyfert galaxies against LINERs. The data sets
used in the experiments and the spectroscopic parametrization used to train the classification methods are described in Section~\ref{sec:data}, while the
detailed description of the different classes of algorithms used is given in Section~\ref{sec:methods}. The results of the experiments and the performances
of the four techniques to each different class of experiment are described in Sec.~\ref{theexperiments}, and we discuss our findings in
Section~\ref{sec:discussion}. Finally, we draw our conclusions in Section~\ref{sec:conclusions}.

\section{The Knowledge Base and the data}\label{thedata}
\label{sec:data}

Supervised methods learn how to reproduce the desired knowledge using the already mentioned Knowledge Base, i.e. a collection of \textit{examples},
that are patterns for which the right classification (target) is already known by independent means. It goes without saying that a biased, incomplete or
poor KB will affect the classification efficiency. In other words, since one of the main drawbacks of the machine learning methods is the difficulty in
extrapolating to regions of the input parameter space that are not well sampled by the training data, the KB needs to cover in a homogeneous way the whole
parameter space with a local density which depends on the complexity of the knowledge to be reproduced.

For the reasons outlined in the previous paragraph, it is quite evident that, in the case of AGNs, the construction of a complete and unbiased KB is almost
an impossible dream unless very conservative choices, such as those adopted in this paper, are made.

Our KB was obtained by merging two different samples (respectively, \citealt{sorrentino2006} and \citealt{kauffmann2003}), of objects for which a
classification based on spectroscopy, was available.

Both samples were drawn from the SDSS DR4 \emph{PhotoSpecAll} table which contains all objects for which both photometric and spectroscopic
observations are available.\\

\noindent \underline{Catalogue by \cite{sorrentino2006}}
This catalogue contains objects in the redshift range ($0.05<z<0.095$). It  provides a classification as Type $1$ (Seyfert I and LINER I), Type $2$
(Seyfert II and LINER II) and non-AGN for $24293$ objects. The data were extracted from the Sloan Digital Sky Survey Data Release $4$ \citep{adelman2006},
and the selection was performed using the traditional approach based on the equivalent width of specific emission lines. In particular, objects classification was originally performed by Sorrentino et al. which assumed to be bona fide AGN sources that lay above one of the so called Kewley's lines, \citep{kewley2001}:
\begin{equation}\label{kewley1}
\log \frac{{[OIII]\lambda 5007}}{{H_\beta  }} = \frac{{0.61}}{{\log
\frac{{[NII]\lambda 6583}}{{H_\alpha  }} - 0.47}} + 1.19
\end{equation}

\begin{equation}\label{kewley2}
\log \frac{{[OIII]\lambda 5007}}{{H_\beta  }} = \frac{{0.72}}{{\log
\frac{{[SII]\lambda \lambda 6717,6731}}{{H_\alpha  }} - 0.32}} +
1.30
\end{equation}

\begin{equation}\label{kewley3}
\log \frac{{[OIII]\lambda 5007}}{{H_\beta  }} = \frac{{0.73}}{{\log
\frac{{[OI]\lambda 6300}}{{H_\alpha  }} - 0.59}} + 1.33
\end{equation}
Furthermore, AGNs were classified as Seyfert I if:

\begin{equation}\label{classificazione1} FWHM(H_\alpha)
> 1.5 FWHM([OIII]\lambda 5007)
\end{equation}

\noindent or

\begin{equation}\label{classificazione2} FWHM(H_\alpha)
>1200 Km s^{-1}\end{equation}

\noindent and

\begin{equation}\label{classificazione3}  FWHM([OIII]\lambda 5007) < 800 Km s^{-1}\end{equation}

\noindent All the other AGNs were classified as Seyfert II. The final catalogue comprises $22464$ objects recognized as non-AGN, $725$ Seyfert I, and
$1105$ Seyfert II (see their figure 2 and 3).\\

\noindent \underline{Catalogue by \cite{kauffmann2003}}:
This catalogue\footnote{\url{http://www.mpa-garching.mpg.de/SDSS/DR4/}} contains spectra lines and ratio for 88178 galaxies ($0.02<z<0.3$). Since this is a purely spectroscopic catalogue, in order to divide objects in different classes, we followed \cite{kauffmann2003} prescriptions, defining a region populated by AGNs above the Kewley's line, (eq. \ref{kewley1}), \citep{kewley2001}, and a second region populated by objects which are likely not to be AGN below the Kauffmann's line \citep{kauffmann2003,kewley2006}:

\begin{equation}\label{kauffmann}
\log \frac{{[OIII]\lambda 5007}}{{H_\beta  }} = \frac{{0.61}}{{\log
\frac{{[NII]\lambda 6583}}{{H_\alpha  }} - 0.05}} + 1.3
\end{equation}

\noindent The intermediate region is heavily contaminated by non-AGN and, in what follows, we shall refer to it as the \textit{mixed} zone.

\noindent Finally, as in the previous case, we used the following Heckman's line \citep{heckman1980,kewley2006},
\begin{equation}\label{LINERS}
    \frac{{[OIII]\lambda 5007}}{{H_\beta  }}=2.1445 \frac{{[NII]\lambda 6583}}{{H_\alpha  }} + 0.465
\end{equation} to further divide the sample in Seyferts and LINERs.

\begin{figure*}
   \centering
   \resizebox{\hsize}{!}{\includegraphics{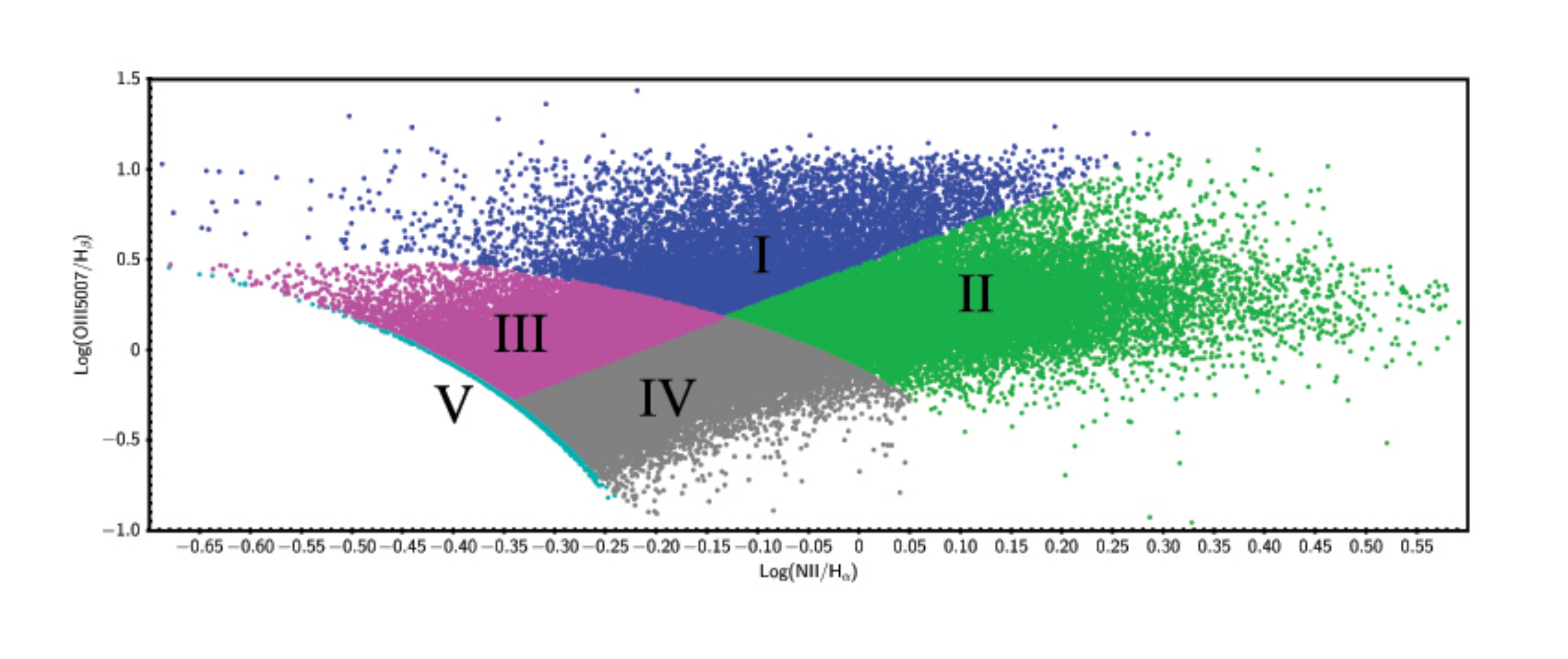}}
     \caption{A representation of the Kauffmann catalogue in the BPT diagram. The blue (I) region is populated by
     Seyferts; green region (II) are LINERs; the violet (III) by a mixture of Seyferts and non-AGN; the grey region (IV) is populated by a mixture of
     non-AGN and LINERs; finally
     light blue region (V) at the lower left boundary is populated by non-AGN.}
\label{Fig:bpt}
\end{figure*}

The resulting five areas in the plane defined by the equivalent widths line ratios (Fig.~\ref{Fig:bpt}), are populated by objects from the Kauffmann sample
classified according to the Baldwin, Phillips and Terlevich (BPT; \citealt{baldwin1981}), diagram diagnostics. In the diagram it is evident that
this catalogue contains few object below the Kauffmann line (surely non-AGN), although other non-AGN objects are present in the mixed zone and in the
catalogue by \cite{sorrentino2006}.\\

\noindent \underline{Catalogue by \cite{dabrusco2007}}:
This catalogue\footnote{\url{http://dame.dsf.unina.it/catalogues.php}} contains photometric redshifts for all SDSS-DR4 with $z < 0.4$, matching
the following selection criteria: dereddened magnitude in $r$ band, $r < 21$; $mode = 1$ which corresponds to primary objects only in the case of de-blended sources.

The first two catalogues were merged together and, in the case of overlapping entries, we retained the Type 1 and Type 2 information from \cite{sorrentino2006}
and all other types from \cite{kauffmann2003}. The resulting merged catalogue included a total of $108162$ objects. The two catalogs have $\sim2000$ common objects. In principle it could be possible to perform the same classification done by \cite{sorrentino2006} on the \cite{kauffmann2003} catalog. The choice has been driven by considering their divergent and specific goal: \cite{sorrentino2006} tried to investigate the AGN environment, by following a very conservative approach; on the other side, the prescriptions of \cite{kauffmann2003} and \cite{kewley2006} are more general-purpose.
It was finally cross matched with the third catalogue, containing photometric redshifts (photo-z) provided by \cite{dabrusco2007}, having a very low standard deviation of residual error ($\sigma \simeq 0.02$), which further reduced the number of objects to 100.069. The need for photo-z was dictated by our goal to identify potential AGN objects using only the available photometric information (a choice which ruled out the use of more accurate spectroscopic redshifts).
For these objects we extracted the following parameters from the SDSS archive and for each band (see SDSS web site\footnote{\url{http://www.sdss.org/}} for
details):

\begin{itemize}
\item $fiberMag$, flux in 3 arcsec diameter fiber radius;
\item $petroMag$, petrosian flux;
\item $petroR50$, radius containing 50\% of petrosian flux;
\item $petroR90$, radius containing 90\% of petrosian flux;
\item $dered$, deredenned magnitude, corrected for extinction.
\end{itemize}

Therefore the number of initial parameters is $26$ (i.e. the five SDSS parameters listed above for each of the five SDSS $ugriz$ bands plus the photometric
redshift). We also used derived parameters such as colors and concentration index. Finally, all objects with undefined values for some of their parameters
(named also as \textit{Not a Number} or NaN) were removed and this reduced the final number of objects to 84885. In this case with the term
\textit{undefined values} we mean undefined numerical values underlying either non detection or contaminated measurements. This last step is crucial in
machine learning methods since the presence of such unknown data might affect their generalization
capabilities \citep{marlin2008}.

\begin{table*}
\begin{center}
\begin{tabular}{@{}|c|c|c|c|c|@{}} \hline
CLASS & CATALOGUE & Exp. AGN vs non-AGN & Exp. Seyfert I vs Seyfert II & Exp. Seyfert vs LINER \\ \hline \hline
Non-AGN & All & Class 0 & - & - \\ \hline
Type 1 & Sorrentino & Class 1 & Class 1 & - \\ \hline
Type 2 & Sorrentino & Class 1 & Class 0 & - \\ \hline
Mix-LINER & Kauffmann & Class 0 & - & - \\ \hline
Mix-Seyfert & Kauffmann & Class 0 & - & - \\ \hline
Pure-LINER & Kauffmann & Class 1 & - & Class 0 \\ \hline
Pure-Seyfert & Kauffmann & Class 1 & - & Class 1 \\ \hline
Mix-LINER-Type1 & overlap & Class 0 & - & - \\ \hline
Mix-Seyfert-Type1 & overlap & Class 0 & - & - \\ \hline
Pure-LINER-Type1 & overlap & Class 1 & Class 1 & Class 0 \\ \hline
Pure-Seyfert-Type1 & overlap & Class 1 & Class 1 & Class 1 \\ \hline
Mix-LINER-Type2 & overlap & Class 0 & - & - \\ \hline
Mix-Seyfert-Type2 & overlap & Class 0 & - & - \\ \hline
Pure-LINER-Type2 & overlap & Class 1 & Class 0 & Class 0 \\ \hline
Pure-Seyfert-Type2 & overlap & Class 1 & Class 0 & Class 1 \\ \hline
SIZE: 24293 & Sorrentino  & 84885 & 1570 & 30380 \\
SIZE: 88178 & Kauffmann  &  &  &  \\ \hline
 \end{tabular}
\end{center}
\caption{The final data set (knowledge base) composition. Empty fields stand for the unused typology. The separation between class $0$ and class $1$ are
referred to the target vector (used during training).}\label{tabledataset}
\end{table*}

This final catalogue, summarized in Tab.~\ref{tabledataset}, was then used to create three different data sets to be used for the three distinct
classification experiments described in Sec.~\ref{theexperiments}. Namely:
\begin{enumerate}
  \item KB data set for the AGN vs non-AGN experiment: the whole (Kauffmann + Sorrentino) catalogue;
  \item KB data set for the Seyfert I vs Seyfert II experiment: just the pure AGN objects belonging to the data set of \cite{sorrentino2006}, resulting
      into 1570
      objects;
  \item KB data set for the Seyferts vs LINERs experiment: pure AGN objects, belonging to the catalogue of \cite{kauffmann2003}, divided into LINERs and
      Seyferts, obtaining 30380 objects.
\end{enumerate}

\section{The methods}\label{themethods}
\label{sec:methods}
DAMEWARE, (cf. \citealt{brescia2010}), is among the main products made available through the DAME
(Data Mining \& Exploration) Program Collaboration.
It provides a web application, able to configure and execute data mining experiments through machine learning
models on a distributed computing infrastructure.
More recently some of these machine learning algorithms were offered in a parallel version which exploits the
computing possibilities offered by Graphical Processing Unit (GPU) technology \citep{cavuoti2013}.\\
From the models available in DAMEWARE we selected four supervised classifiers, the Support Vector Machines
(SVM) and three variants of the Multi Layer Perceptron (MLP), a standard neural network trained by different types of
self-adaptive learning rules, respectively, the Coniugate Gradient (CG), the Scaled Conjugate Gradient (SCG) and the
Quasi Newton Algorithm (MLPQNA).

\underline{Support Vector Machines} \citep{chang2001}:  SVM are supervised learning models with associated
learning algorithms that analyze data and recognize patterns, which are mostly used for classification and regression
analysis. The basic SVM takes a set of input data and predicts, for each given input, which of two possible classes
forms the output, making it a non-probabilistic binary linear classifier. Given a set of training examples, each marked
as belonging to one of two categories, an SVM training algorithm builds a model that assigns new examples into one
category or the other. An SVM model is a representation of the examples as points in space, mapped so that the
examples of the separate categories are divided by a clear gap that is as wide as possible. New examples are then
mapped into that same space and assigned to a category depending on which side of the gap they fall on. In
addition to performing linear classification, SVMs can efficiently perform non-linear classification using what is called
the kernel trick, implicitly mapping their inputs into high-dimensional feature spaces. Among different types of
such kernel function, we used the radial basis function type \citep{chang2001}.

The SVM training experiments over large data sets have huge computational cost (about one week per experiment on a single CPU for a data sample of about
$80000$ input patterns), thus, in order to be able to perform the hundreds of experiments described in what follows (see Sec.~\ref{theexperiments}), it
was needed to exploit the SCoPE\footnote{\url{http://www.scope.unina.it/C19/astrophysics-gridcomputing}} GRID infrastructure resources
\citep{brescia2009}.

\underline{Multi Layer Perceptron} \citep{bishop1995}:  Neural Networks (NNs) have long been known to be excellent tools for interpolating data and for
extracting patterns and trends and since a few years they have also carved  their way into the astronomical community for a variety of applications (see
the reviews
\citealt{tagliaferri2003a,tagliaferri2003b} and references there in), ranging from star-galaxy separation \citep{donalek2006}, spectral classification
\citep{winter2004}, and photometric redshifts evaluation \citep{cavuoti2012,brescia2013}.
In practice a neural network is a tool which takes a set of input values (input neurons), applies a non-linear (and unknown) transformation and returns an
output. The optimization of the output is performed by using a set of examples for which the output value (target) is known a priori. Performances of a MLP
are
greatly affected by the choice of the learning rule, i.e. by the mathematical expression used for the optimization of its internal weights. In this paper
we tested three different
rules, namely, the Coniugate Gradient (CG), the Scaled Conjugate Gradient (SCG), and the Quasi Newton Algorithm (MLPQNA).
In essence, the learning process of a MLP consists of two phases through the different layers of the network: a forward
pass and a backward pass. In the forward pass, an input vector is applied to the input nodes of the network, and its
effect propagates through the network layer by layer. Finally, a set of outputs is produced as the actual response of
the network. During the backward pass, on the other hand, the weights are all adjusted in accordance with the error-
correction rule.
The training of NNs like MLP implies to find the more efficient among a population of NNs differing in the hyper-
parameters controlling the learning of the network, in the number of hidden nodes, etc. The most important hyper-
parameter (usually called $\alpha$), is related to the weights of the network and allows to estimate the dependency of
the training performance on the different inputs and the selection of the parameters for a given task. In fact, a larger
value of $\alpha$ implies a less meaningful corresponding weight \citep{bishop1995}. The three variants of learning
rules discussed here differ basically in the way to calculate the $\alpha$ parameter.

All these algorithms found the minimum of a square error function, but the computational cost of each step is high,
because in order to determine the values of $\alpha$, we have to refer to the \emph{Hessian matrix} $H$ of the
error, which is highly expensive in terms of calculations. But fortunately, the coefficients like the parameter $\alpha$ can be
obtained from analytical expressions that do not use the Hessian matrix explicitly. The method of Conjugate
Gradients reduces the number of steps to minimize the error up to a maximum of $|w|$ (where $|w|$ is the
cardinality of network weights), because there could be almost $|w|$ conjugate directions in a $|w|$-dimensional
space \citep{golub1999}. The Scaled Conjugate Gradients method differs from the CG by imposing that the
Hessian matrix $H$ is always positive \citep{nocedal1999}. This can be done by adding to $H$ a multiple of
identity matrix $\lambda I$, where $I$ is the identity matrix and $\lambda>0$ is a scaling
coefficient \citep{watrous1987}. Finally the Quasi Newton Algorithm does not calculate the $H$ matrix, but an
approximation
in a series of steps. A famous implementation of the QNA, which offers good performance even for non-smooth
optimizations, is known as BFGS, by the names of its inventors \citep{broyden1970,fletcher1970,goldfarb1970,shanno1970}, and it was our choice. This
approach generates a
sequence of matrices $G$ which are subsequent more and more accurate approximations of the Hessian matrix by using only information
related to the first derivative of error function \citep{brescia2012}.

In what follows we outline the standard data mining procedure which was adopted in all the AGN classification experiments performed with the machine
learning models
described above.

\subsection{Feature extraction}
The first step is the pruning of the input parameters. Most machine learning methods are in fact quite demanding in terms of computing time, which may
scale badly with the number of input parameters (features). It is therefore necessary to optimize the number of input features by performing what is
usually called the \emph{feature selection or pruning} phase, aimed at
identifying the subset of features carrying the highest amount of information for a specific task.\\
In order to perform the pruning of input features, the initial $26$ features have been organized by replacing the five magnitudes $dered$ for each band
with the corresponding colors plus the $r$ $dered$ magnitude as reference, due to their capability to improve the performance, as revealed after some
preliminary experiments. The improvement carried by colors can be easily understood by noticing that even though colors are derived as a subtraction of
magnitudes, the content of information is quite different, since an ordering relationship is implicitly assumed, thus increasing the amount of information
in the final output (gradients instead of fluxes). The additional reference magnitude instead removes the degeneracy in the luminosity class for a specific
galaxy type \citep{brescia2013}.

Then a \textit{leave-one-out} cyclic method has been used to test the contribute of each single feature to the classification training performance and to
remove each time the worst resulting. This cyclic procedure is stopped when the performance does not increase by removing any further feature.\\

The leave-one-out procedure was performed according to the following top-down strategy:
\begin{enumerate} \item perform one experiment with all the features and store the performance;
\item perform $N$ experiments, where $N$ is the whole number of features in the data set, by removing each time one of the features;
\item find the set of features achieving the best performance;
\item if the achieved performance is better or equal than the previous one, remove the feature from the set, store the result and go back to point ii,
    otherwise stop the procedure.
\end{enumerate}

At the end the feature extraction phase produced the following subset of $7$ selected input features, candidates to perform the final classification
experiments:\\
\begin{itemize}
\item the $4$ SDSS colors $(u-g), (g-r), (r-i), (i-z)$, deredenned for galactic absorption;
\item the deredenned magnitude in the $r$ band;
\item $fibermag\_r$ the fiber magnitude in the $r$ band;
\item the photometric redshift derived from \cite{dabrusco2007}.
\end{itemize}

All these features have been used in all the experiments in order to maintain the coherence along the overall classification process.

\subsection{Model architecture selection}
The second step consists in identifying for each model, via a trial-and-error procedure, the best architecture, which, for instance, in the case of MLP
would mean to find the optimal number of neurons in the hidden layer, the optimal learning function, etc. Since there is no way to define it \textit{a
priori}, it is necessary to perform many experiments changing every time the parameters defining the model.

In each experiment, the KB is randomly split into two parts, namely the training set ($70\%$ of the data set) to be used by the model to learn the classification
rule and the test set (the remaining $30\%$), used exclusively to evaluate the results.
Due to the supervised nature of the classification task, the system performance can be measured by means of a test
set during the testing procedure, in which unseen data are given to the system to be labeled. The overall performance thus integrates information about the classification accuracy (i.e. in terms of output correctness). Moreover, the results obtained from the unseen data are also important to evaluate the learning robustness, i.e. the generalization capability of the network in presence of data samples never used during the training phase.\\
Furthermore, it is important to stress that, in order to ensure a proper coverage of the KB in the Parameter Space (PS), the data objects were indeed divided up among the training and blind test datasets by random extraction, and this process minimizes the possible biases induced by statistical fluctuations in the coverage of the PS, namely small differences in the class distribution of training and test samples used in the experiments.

\subsection{Evaluation of performances}
Performances were evaluated on the test set using a standard set of statistical indicators defined in this section. We wish to stress that the test were
performed by submitting to any given model the photometric data alone and then by comparing the predicted value with the target. For a given confusion
matrix:
\begin{equation}
  \begin{array}{c|ccc}
            &            & $OUTPUT$\\ \hline
            &    -       &$Class A$      & $Class B$ \\
    $TARGET$  &$Class A$    & N_{AA}              & N_{AB} \\
            &$Class B$    & N_{BA}              & N_{BB} \\
  \end{array}
  \end{equation}

We can define the following statistical quantities:

\begin{itemize}
\item \underline{total efficiency}: $te$. Defined as the ratio between the number of correctly classified objects and the total number of objects in the
    data set. In our confusion matrix example it would be:
    \begin{equation}
    te=\frac{N_{AA} + N_{BB}}{N_{AA} + N_{AB} + N_{BA} + N_{BB}}
    \end{equation}
\item \underline{purity of a class}: $pcN$. Defined as the ratio between the number of correctly classified objects of a class and the number of objects
    classified in that class, also known as efficiency of a class. In our confusion matrix example it would be:
    \begin{equation}
    pcA=\frac{N_{AA}}{N_{AA}+N_{BA}}\\
    \end{equation}
    \begin{equation}
    pcB=\frac{N_{BB}}{N_{AB}+N_{BB}}\\
    \end{equation}
\item \underline{completeness of a class}: $cmpN$. Defined as the ratio between the number of correctly classified objects in that class and the total
    number of objects of that class in the data set. In our confusion matrix example it would be:
    \begin{equation}
    cmpA=\frac{N_{AA}}{N_{AA}+N_{AB}}\\
    \end{equation}
    \begin{equation}
    cmpB=\frac{N_{BB}}{N_{BA}+N_{BB}}\\
    \end{equation}
\item \underline{contamination of a class}: $cntN$. It is the dual of the purity, namely it is the ratio of misclassified object in a class and the
    number of objects classified in that class, in our confusion matrix example will be:
    \begin{equation}
    cntA=1-pcA=\frac{N_{BA}}{N_{AA}+N_{BA}}\\
    \end{equation}
    \begin{equation}
    cntB=1-pcB=\frac{N_{AB}}{N_{AB}+N_{BB}}\\
    \end{equation}
\end{itemize}

\section{Experiments and results}\label{theexperiments}
We performed three different kinds of experiments: (i) AGN detection; (ii) Seyfert I vs Seyfert II classification and (iii) Seyferts vs LINERs
classification. In all cases the experiments were approached with two kinds of machine learning models, respectively, SVM and MLP, the latter in three
different versions, by changing the internal learning rule (i.e. CG, SCG and QNA), as described in Sec.~\ref{themethods}.

\subsection{AGN classification}
Concerning the classification of AGN against non-AGN, the MLP models were trained  using a target vector
whose values where set to 1 for each object above the Kewley's line (i.e. pure AGN) and to 0 for object below it (which therefore includes the mixed zone
objects which are non-AGN). The KB included 84885 objects after the removal of the patterns affected by NaNs. According to the mentioned strategy, the
training set ($70\%$ of the whole data set) contained $59419$ patterns while the test set ($30\%$ of the data set) contained $25466$ patterns.

The MLP output may be interpreted as the probability for a given object to belong to a specific class and a threshold needs to be assumed in order to classify the objects. With the standard choice of such threshold to $0.5$, for instance, an object above the threshold is considered to belong to the class of AGNs. Such threshold represents the median point of the probability to assign the MLP output to a class or another in a two-class classification problem.
As it can be seen from the Tab.~\ref{riepilogoAGN}, which summarizes the results outcoming from all the experiments, the best result was obtained by the MLP with the Quasi Newton learning rule.

\begin{table}
  \centering
  $\begin{array}{|c|c|c|c|c|c|} \hline
$model$	& $te$ 	    & $cmpAGN$  & $cmpMIX$   & $pcAGN$	& $cntAGN$  \\ \hline \hline
$CG$	& $75.5\%$	& $55.6\%$  & $86.3\%$	 & $68.5\%$	& $31.5\%$ \\ \hline
$SCG$	& $75.7\%$	& $55.1\%$  & $86.2\%$	 & $68.4\%$	& $31.6\%$ \\ \hline
$QNA$	& $76.5\%$	& $58.6\%$  & $86.5\%$	 & $70.8\%$	& $29.2\%$ \\ \hline
$SVM$   & $75.8\%$  & $55.4\%$  & $86.2\%$   & $70.6\%$ & $29.4\%$ \\ \hline
\end{array}$
 \caption{AGN vs non-AGN: the first column is the model used, while the others give, respectively, the total efficiency, the AGN completeness, the non-AGN
 completeness, the AGN purity and the AGN contamination. These percentages are calculated by considering only the results on the $25466$ objects of
 test set (i.e. not including training set results).}\label{riepilogoAGN}
\end{table}

In the best experiment, the SVM reached a comparable result, $75.8\%$,  obtained with  $C = 32768$ and $\gamma = 0.001953125$, where $C$ is a penalty
parameter and $\gamma$ the internal parameter of the radial basis function kernel \citep{chang2001}. The Tab.~\ref{riepilogoAGN} reports the complete
results by using the three MLP learning rules and the SVM.

\begin{table}
  \centering
  $\begin{array}{|c|c|c|c|c|} \hline
$excluded$	        & $te$         & $cmp$       & $pc$        & $cnt$       \\
$features$          &              &             &             &             \\ \hline \hline
$photo-z$	        & $-0.4\%$     & $-0.8\%$    & $-0.6\%$	   & $+0.6\%$    \\ \hline
$fibermag\_r$       & $-0.9\%$	   & $-2.8\%$    & $-1.7\%$    & $+1.7\%$    \\ \hline
$dered\_r$	        & $-0.6\%$     & $-2.5\%$    & $-0.3\%$    & $+0.3\%$    \\ \hline
$all except colors$ & $-0.9\%$     & $-1.6\%$    & $-3.7\%$    & $+3.7\%$    \\ \hline
\end{array}$
 \caption{AGN vs non-AGN: comparison among the $7$ features of the test set in terms of their amount of information given to the classification
 performance. The first column reports the excluded features from the training set for each pruning experiment. The others are respectively, total
 efficiency, completeness, purity and contamination. The values are expressed in terms of percentage variations in respect to the best values (QNA)
 reported in Tab.~\ref{riepilogoAGN} }\label{newpruning}
\end{table}

\subsection{Seyfert I vs Seyfert II classification}
In the classification between type I and type II Seyfert objects, the ML models were fed using a target vector whose values were set to $1$ for objects
classified as Seyfert I in the catalogue by \cite{sorrentino2006} and to $0$ if classified as Seyfert II, resulting in $1830$ objects and $1570$ after the
usual removal of the patterns affected by NaN values. So, the training set contained $1256$  patterns while the test set $314$ patterns.

In this case, the main parameter of interest that quantifies the ability to distinguish the two classes is the efficiency; the MLPQNA model produced a total efficiency of $72\%$, while using the SVM, the best result produced a total efficiency equal to $81.5\%$. As it
can be seen the results are promising, even more if we take into account the small number of patterns used for the training.

\subsection{Classification of Seyferts vs LINERs}

Concerning the last experiment, namely the classification between Seyfert and LINER objects, the ML models were fed using a target vector with values
labeled as $1$ for objects laying below the Heckman's line, and $0$ for objects above the line. This resulted in a total of $30380$ objects after the
removal of the patterns affected by NaN presence.
The training set ($70\%$ of the whole data set) contained $21266$  patterns and the test set $9114$ patterns.

The MLPQNA model produced a total efficiency of $73.8\%$, while using SVM we reached the best total efficiency equal to $78.18\%$.

\section{Discussion of the experiments}
\label{sec:discussion}

In general terms, in the main experiment (i.e. the classification AGN vs non-AGN), the MLP with QNA learning rule performs better than all other methods,
both in terms of performance and robustness. This is not completely a surprise since already in other cases, the MLPQNA has been proven \citep{brescia2012}
to be quite effective in optimizing the poor information introduced by a small or incomplete KB, due to its fine approximation of the Hessian of the
training error \citep{broyden1970,fletcher1970,goldfarb1970,shanno1970}.
We obtain a good overall efficiency, of about $75\%$ with a good purity ($70\%$) while all methods performed badly in terms of completeness reaching about $58\%$ even though it must be stressed that if a high level of purity is needed for specific applications, MLPQNA can be fine tuned to do it by varying the threshold at which an object is recognized as AGN, but this can be done at the price of a loss in
completeness. The low completeness may be partly explained by the ambiguities introduced by template patterns in the mixed zone. We therefore investigated the possibility to increase the purity at a relative price of completeness by changing the threshold level. The optimal value has been obtained for the threshold $0.87$ which leads to
a purity of $88\%$ and a level of completeness of $9\%$. It goes without saying that the balancing between purity and completeness can be performed according to the needed from the specific application.\\

By considering the data set which gives the best results, obtained with the MLPQNA model, we performed a series of experiments to evaluate the
contribution of each feature of training objects to the test performance, in terms of information given to the classification during training. This set of
tests has been done by alternately excluding some of the features for all training objects. The resulting variation percentages for all used statistical
indicators are shown in Tab.~\ref{newpruning}. We emphasize that in our case photometric redshifts are crucial to reproduce the same cut at spectroscopical redshift $<0.3$, imposed by the original knowledge base \citep{kauffmann2003}. This is a typical requirement of empirical methods, in order to maintain the coherence between trained and new data samples in terms of parameter space.\\

Concerning the analysis of the contribution to the classification performance of the photometric features, composing the training and test patterns, the
series of tests, reported in Tab.~\ref{newpruning}, have shown a significant valence of colors and reference magnitudes (mainly fibermag but also dered in
r band), followed by an important contribution of photometric redshift. Although not surprising for colors, due to their objective quantity of correlated
information carried, it resulted quite interesting that without information given by photo-z and reference magnitudes, the classification capability
underwent a significant decrease.\\

By considering the subset of non-AGN objects within the class including both mixed and non-AGN objects, its percentage of false positives (i.e.
those misclassified as AGN) is about $1\%$. Moreover the percentage of objects, spectroscopically known as non-AGN, which become false positive is also
about $1\%$. The contamination due to galaxies is very small, and this must be considered very encouraging since the ambiguities in the knowledge base,
introduced by unrecognized AGN in the mixed zone, can only lower this already very small percentage.

Concerning the experiment related to the classification of objects in Seyfert I vs Seyfert II (hereinafter experiment $2$), the level of performance can be
easily understood in terms of the small dimension of the training data set, since in general Machine Learning techniques are quite sensitive to the
incompleteness of the KB.\\

About the classification Seyfert vs LINER (hereinafter experiment $3$), the contamination in the lower region near to the Heckman's line \emph{confuses}
the machine learning techniques, leading, in turn, to reduced performances of the photometric classification. This is also partially true in the first
experiment (AGN vs non-AGN), where a contamination is also present in the mixed zone, i.e. between Kauffmann and Kewley lines.\\

Hence, a clear result of our experiments is that an unambiguous KB is required to successfully train and apply any classification method. This can be
brought back to the fact that Seyfert I and Seyfert II, from the optical photometry point of view, show substantially different behavior, while the
difference between Seyfert and LINER is somehow more vague; this situation is even worse in the AGN vs non-AGN experiment where the whole mixed zone
\emph{confuses} the network. This is also evident in the spectroscopic parameter space where the so called \emph{seagull wings} move away far from the
Kewley line. By considering the Seyfert I and II alone, it results evident a quite sharp spectroscopical separation \citep{sorrentino2006}.

\section{Conclusions}
\label{sec:conclusions}

The production of large and accurate AGN catalogues is an important topic that will become crucial with the advent of the future photometric only
digital surveys that will map large fractions of the sky to unprecedented depth in the different wavelengths.\\

We have applied four distinct classification methods, based on self-adaptive classification techniques, to the problem of the classification of
emission line galaxies using only optical photometric parameters. The methods have been applied to three classification problems, specifically the
separation of AGNs from non-AGNs, Seyfert I from Seyfert II and the classification of Seyfert from LINERs. In terms of classification efficiency, the results indicate that our methods perform fairly ($\sim76.5\%$) when applied to the problem of the classification of AGNs vs non-AGNs, while the performances in the more fine classification of Seyfert vs LINERs are $\sim78\%$ and $\sim81\%$ in the case Seyfert I vs Seyfert II.\\

From a methodological standpoint, the results of our experiments indicate how sensitive the performances of the photometric classification of line-emission
galaxies are to the size of the spectroscopic data sets used to train the method, and to the uncertainty affecting the spectroscopic classification of the
training set sources.\\

It is important to stress that, even with a completeness of about $58\%$, the possibility to use photometric data alone would led to a catalogue of
candidate AGN about $200$ times larger than existing ones, still retaining a purity of about $70\%$.\\
This work, that should be interpreted as a \emph{feasibility study}, is hence just a first step and encourages the possibility to proceed further with more
fine classifications of the different families of line emission galaxies by exploiting their multi-band photometry.

\section*{Acknowledgments}

\noindent The authors would like to thank the anonymous referee for the comments and suggestions which helped us to improve the paper.
\noindent The authors wish to thank the whole DAMEWARE working group, whose huge efforts made the DM facility available to the scientific community.
\noindent MB wishes to thank the financial support of PRIN-INAF 2010, \textit{Architecture and Tomography of Galaxy Clusters}.
\noindent The authors also wish to thank the financial support of Project F.A.R.O. III Tornata (P.I. Dr. M. Paolillo, University Federico II of Naples).
\noindent GL acknowledges financial contribution through the PRIN-MIUR 2012 Euclid.

\end{document}